\documentclass[twocolumn,prl,aps,superbib,tightenlines,floatfix,superscriptaddress,showpacs]{revtex4}
\usepackage[T1]{fontenc}
\usepackage[latin9]{inputenc}
\usepackage{amsmath}
\usepackage{graphicx}
\usepackage{amssymb}

\begin{document}

\title{Spin relaxation in Rashba rings}

\author{Valeriy A. Slipko}
\affiliation{Department of Physics and Astronomy and USC
Nanocenter, University of South Carolina, Columbia, SC 29208, USA}
\affiliation{ Department of Physics and Technology, V. N. Karazin
Kharkov National University, Kharkov 61077, Ukraine }

\author{Yuriy V. Pershin}
\email{pershin@physics.sc.edu}

\affiliation{Department of Physics and Astronomy and USC
Nanocenter, University of South Carolina, Columbia, SC 29208, USA}

\begin{abstract}
Spin relaxation dynamics in rings with Rashba spin-orbit coupling is investigated using spin kinetic equation.
We find that the spin relaxation in rings occurs toward a persistent spin configuration whose final shape depends on the initial spin polarization profile. As an example, it is shown that a homogeneous parallel to the ring axis spin polarization transforms into a persistent crown-like spin structure. It is demonstrated that the ring geometry introduces a geometrical contribution to the spin relaxation rate speeding up the transient dynamics. Moreover, we identify several persistent spin configurations as well as calculate the Green function of spin kinetic equation.
\end{abstract}

\maketitle

\section{Introduction}

The understanding of spin relaxation dynamics in semiconductor structures is an active area of research related to the field of spintronics \cite{Zutic04a,Wu10a}. Previously, the spin relaxation dynamics has been investigated mainly in infinite two-dimensional (2D) systems with D'yakonov-Perel' \cite{Dyakonov72a,Dyakonov86a} spin relaxation mechanism (see, e.g., Refs. \cite{Dyakonov86a,Sherman03a,Pershin05a,Bernevig06a,Weng08a,Kleinert09a,Tokatly10a,pershin10a}). There are only several examples in the literature
where the effects of boundary conditions \cite{Galitski06a} on D'yakonov-Perel' spin relaxation have been explored. These examples include investigations of spin relaxation in 2D channels \cite{Kiselev00a,Holleitner07a}, 2D half-space \cite{Pershin05c}, 2D systems with antidots \cite{Pershin04a}, small 2D systems \cite{Lyubinskiy06a}, and finite-length one-dimensional (1D) structures \cite{Slipko11a}. The main conclusion of all these studies is that the introduction of boundary conditions results in an increased spin lifetime. In particular, in Ref. \cite{Lyubinskiy06a} Lyubinskiy reports on a strong suppression of spin relaxation in small 2D systems. In Ref. \cite{Slipko11a} the present authors demonstrate that a persistent spin helix spontaneously emerges in course of relaxation of homogeneous spin polarization. As real electronic devices are always of a finite size, the understanding of spin relaxation dynamics in reduced geometries is of a crucial importance.

\begin{figure} [b]
\begin{center}
\includegraphics[angle=0,width=5.0cm]{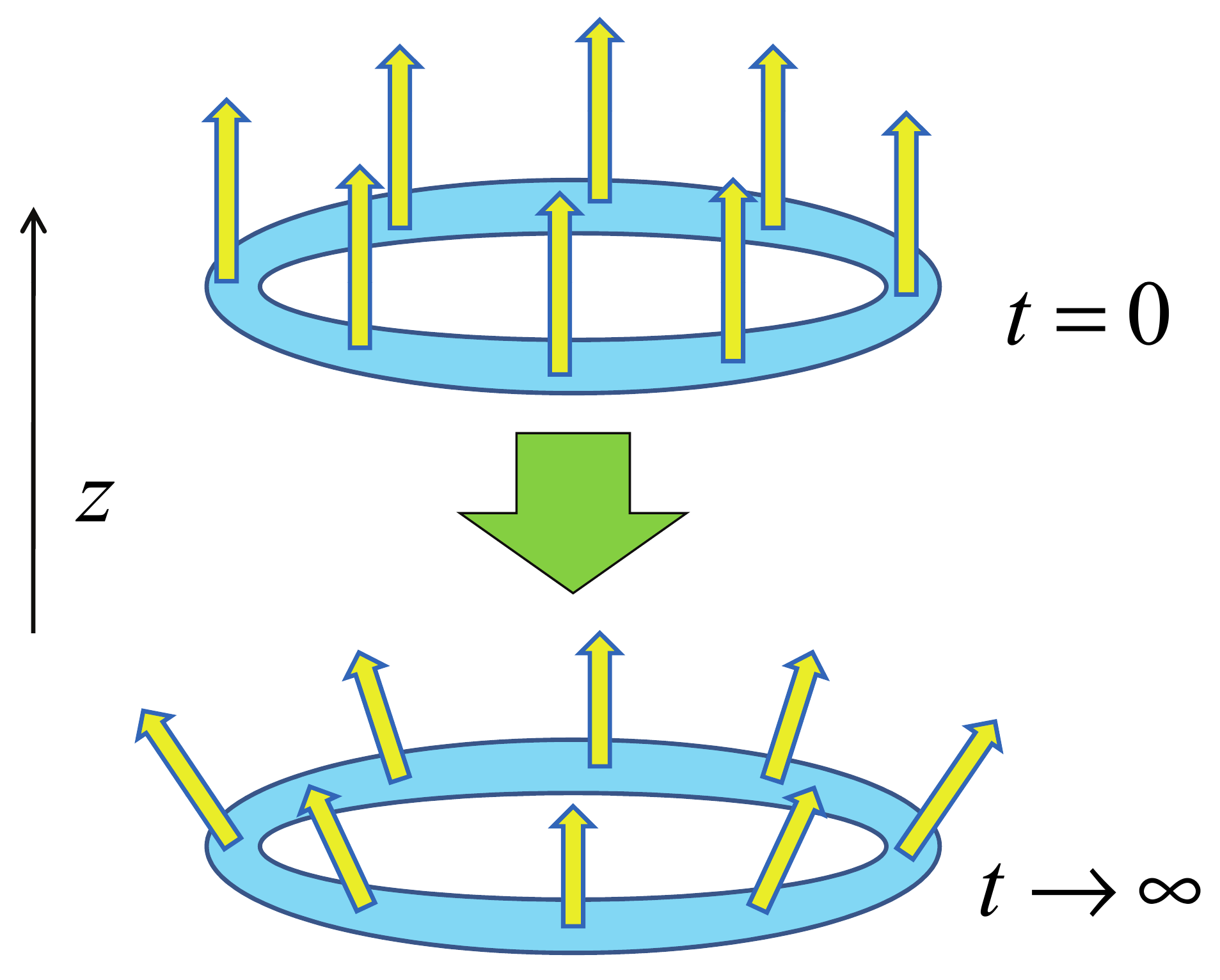}
\caption{(Color online) Schematic diagram of spin relaxation in a ring. Initially homogeneous
spin polarization in $z$ direction (along the ring axis) transforms into a persistent crown-like spin structure. Note that the spin polarization
amplitude decreases in this process according to Eq. (\ref{SzAttenuation}).}
\label{fig1}
\end{center}
\end{figure}

In this paper, we present our studies of spin relaxation in rings with Bychkov-Rashba \cite{Bychkov84a} spin-orbit coupling. The ring geometry is especially interesting since on the one hand, the electron space motion in rings is confined to a limited space region and on the other hand, the ring geometry does not require any boundary conditions (such as those derived in Ref. \cite{Galitski06a}) for 1D transport that normally have to be used, for example, to describe spin dynamics in finite-length structures \cite{Slipko11a}. Starting with a spin kinetic equation, we formulate a set of equations for spin polarization components on the ring and solve these equations in some particular cases. Specifically, we find that a homogeneous spin polarization transforms into a persistent crown-like spin structure schematically shown in Fig. \ref{fig1}. Moreover, we determine a set of non-trivial spin persistent states that are realized at some specific sizes of the ring. Finally, we derive expression for Green function that can be used to find evolution of spin polarization for any initial spin polarization profile. All these results indicate an unusual character of spin relaxation in finite size structures that is dramatically different from the spin relaxation in the bulk.

\section{Spin kinetic equation}
In this section, we introduce the model and derive equations governing spin relaxation dynamics in rings. The main results of this section are given by Eqs. (\ref{Sr})-(\ref{Sz}). The derivation of these equations is based on a kinetic approach. It is important to mention that Eqs. (\ref{Sr})-(\ref{Sz}) are more general than traditional spin drift-diffusion equations \cite{Yu02a,Burkov04a,Saikin04a,Pershin04b,arXiv_1007_0853v1}. In particular, Eqs. (\ref{Sr})-(\ref{Sz}) are applicable in both diffusive and ballistic regimes at any value of spin rotation angle per mean free path.

Let us consider spin-polarized electrons that can move along a ring of a sufficiently large radius $r\gg \hbar/p$ so that small-radius corrections \cite{Meijer02a,Berche10a} to spin-orbit Hamiltonian \cite{Bychkov84a} can be neglected. The spin-orbit Hamiltonian is taken in the form $H_{SO}=\alpha\left(\hat{\mathbf\sigma}\times\mathbf{\hat p}\right)\cdot\mathbf{e}_z$.  Here, $\mathbf{\hat{p}}=(\hat p_x,\hat p_y)$ is the 2D electron momentum operator, $m$ is the effective electron's mass, $\mathbf{\hat\sigma}$ is the Pauli-matrix vector, $\alpha$ is the spin-orbit coupling constant and $\mathbf{e}_z$ is a unit vector along the ring axis. It can be shown that a motion of an electron with a momentum $\mathbf{p}$ along a straight trajectory is accompanied by a spin rotation with the angular velocity $\mathbf{\Omega_p}=\Omega\mathbf{p}\times\mathbf{e}_z/p$, where
$\Omega=2\alpha p/\hbar$. The spin precession angle per unit length is given by $\eta=2\alpha
m \hbar^{-1}$. The ring geometry, however, changes the character of spin rotations since the spin rotation axis rotates as electron moves along the ring.

Our consideration of spin dynamics in rings is based on the kinetic equation for electron spin polarization (see, e.g., Ref. \cite{Lyubinskiy06a,Lyubinskiy06b}).  In quasi-classic approximation such an equation can be written as
\begin{eqnarray}
\left(\frac{\partial }{\partial t}+\frac{\mathbf{p}}{m}\cdot\nabla\right)
\mathbf{S_p}=\mathbf{\Omega_p}\times\mathbf{S_p}+St\{\mathbf{S_p}\},
\label{KinEq}
\end{eqnarray}
where $\mathbf{S_p}(\mathbf{r},t)$ is the vector of spin polarization of electrons, and $St\{\mathbf{S_p}\}$ is the collision integral describing electron scattering processes. Eq. (\ref{KinEq}) describes change in spin polarization of electrons moving with momentum $\mathbf{p}$. The RHS of Eq. (\ref{KinEq}) includes two terms describing rotation of spin polarization with the angular velocity $\mathbf{\Omega_p}$ and change in spin polarization due to scattering processes. In the $\tau$-approximation \cite{Pitaevskii81a} the collision integral is given by
\begin{eqnarray}
St\{\mathbf{S_p}\}=-\frac{1}{\tau}(\mathbf{S_p}-\langle\mathbf{S_p}\rangle),
\label{CollisionIntegral}
\end{eqnarray}
where the angle brackets denote averaging over direction of electron momentum.
The collision integral (\ref{CollisionIntegral}) corresponds to the elastic scattering of electrons by strong scatterers with a characteristic time $\tau$ between the collisions.
For 1D case the average spin polarization simplifies to the following expression $\langle\mathbf{S_p}\rangle=(\mathbf{S}^+ +\mathbf{S}^-)/2 $, where $\mathbf{S}^+$ and $\mathbf{S}^-$ are the spin polarizations of electrons  moving along the ring in the clockwise (with momentum $\mathbf{p}=mv\mathbf{e}_\theta$), and counterclockwise ($\mathbf{p}=-mv\mathbf{e}_\theta$) directions with the average velocity $v$ connected to the mean-free path $\ell$ by $\ell=v \tau$. Thus, the kinetic equation (\ref{KinEq}) for 1D ring of a radius $r$ takes the form of the system of two vector equations
\begin{eqnarray}
\left(\frac{\partial }{\partial t}+\frac{v}{r}\frac{\partial }{\partial \theta}\right)
\mathbf{S}^+=\Omega\mathbf{e}_r\times\mathbf{S}^+ -\frac{1}{2\tau}(\mathbf{S}^+-\mathbf{S}^-),
\label{KinEqPlus}  \\
\left(\frac{\partial }{\partial t}-\frac{v}{r}\frac{\partial }{\partial \theta}\right)
\mathbf{S}^-=-\Omega\mathbf{e}_r\times\mathbf{S}^- -\frac{1}{2\tau}(\mathbf{S}^--\mathbf{S}^+).
\label{KinEqMinus}
\end{eqnarray}
This system of equations should be complimented by the initial conditions for spin polarizations $\mathbf{S}^+(\theta,t=0)$ and $\mathbf{S}^-(\theta,t=0)$ (for clockwise and counterclockwise moving electrons).

Taking the sum and difference of Eqs. (\ref{KinEqPlus}, \ref{KinEqMinus}) we easily obtain
\begin{eqnarray}
\frac{\partial \mathbf{S}}{\partial t}
&=&-\frac{v}{r}\frac{\partial \mathbf{\Delta}}{\partial \theta}
+
\Omega\mathbf{e}_r\times\mathbf{\Delta},
\label{KinEqS}  \\
\frac{\partial \mathbf{\Delta}}{\partial t}
&=&-\frac{v}{r}\frac{\partial \mathbf{S}}{\partial \theta}
+
\Omega\mathbf{e}_r\times\mathbf{S}-\frac{\mathbf{\Delta}}{\tau}
\label{KinEqDelta},
\end{eqnarray}
where the following notations are used: $\mathbf{S}=\mathbf{S}^++\mathbf{S}^-$ and  $\mathbf{\Delta}=\mathbf{S}^+-\mathbf{S}^-$.
As we are mainly interested  in finding the total spin polarization $\mathbf{S}$, $\mathbf{\Delta}$ can be eliminated from Eqs. (\ref{KinEqS}) and  (\ref{KinEqDelta}) via a simple transformation. The final equations for three components of electron spin polarization in cylindrical coordinates, $\mathbf{S}=S_r\mathbf{e}_r+S_\theta\mathbf{e}_\theta+S_z\mathbf{e}_z$,  can be presented in the form
\begin{eqnarray}
\frac{\partial^2 S_r}{\partial t^2}+\frac{1}{\tau}\frac{\partial S_r}{\partial t}=
\omega^2 \frac{\partial^2 S_r}{\partial \theta^2}-\omega^2 S_r-2\omega^2\frac{\partial S_\theta}{\partial \theta}-\omega\Omega S_z, \label{Sr} \\
\frac{\partial^2 S_\theta}{\partial t^2}+\frac{1}{\tau}\frac{\partial S_\theta}{\partial t}=\omega^2 \frac{\partial^2 S_\theta}{\partial \theta^2}-(\omega^2+\Omega^2)S_\theta  \;\;\;\;\;\;\;\;\;\;\;\;\;\;\;\;\;\;\;\;\;    \label{Stheta} \\ \nonumber
+2\omega^2\frac{\partial S_r}{\partial \theta}+
2\omega\Omega \frac{\partial S_z}{\partial \theta},\\
\frac{\partial^2 S_z}{\partial t^2}+\frac{1}{\tau}\frac{\partial S_z}{\partial t}=
\omega^2 \frac{\partial^2 S_z}{\partial \theta^2}-\Omega^2 S_z- \omega\Omega S_r -2\omega\Omega \frac{\partial S_\theta}{\partial \theta},\label{Sz}
\end{eqnarray}
where $\omega=v/r$ is the angular velocity.


The initial rate of change of spin polarization $\mathbf{S}$ may be calculated by using the initial condition for $\Delta$ from Eq. (\ref{KinEqS}). In particular, if polarizations of electrons moving clockwise and counterclockwise at the initial moment are the same, i.e. $\Delta(\theta,t=0)=0$, then we find from Eq.  (\ref{KinEqS}) that in this case
\begin{eqnarray}
\left(\frac{\partial \mathbf{S}}{\partial t}\right)_{t=0}=0.
\label{IC1}
\end{eqnarray}
The diffusive limit Eqs. (\ref{Sr})-(\ref{Sz}) is realized on time scales much longer than $\tau$. In this case, we can neglect $\partial^2 S_i/\partial t^2$ terms in these equations and notice that the resulting equations are of diffusion type with the diffusion coefficient $D=v^2\tau$.

\section{Relaxation of homogeneous spin polarization}
The most interesting and simple solution of Eqs. (\ref{Sr})-(\ref{Sz}) describes the relaxation of the homogeneous spin polarization initially directed along $z$-axis (see Fig.\ref{fig1})
\begin{eqnarray}
\left(\mathbf{S}\right)_{t=0}=S_0\mathbf{e}_z,
\label{IC2}
\end{eqnarray}
where $S_0$ is the initial spin polarization amplitude. Since the initial condition (\ref{IC2}) is symmetrical with respect to rotations about the ring axis,  $S_\theta=0$ at anytime, and the spin polarization components $S_r$ and $S_z$ should not depend on $\theta$. Then Eq. (\ref{Stheta})
satisfies identically, and Eqs. (\ref{Sr}) and (\ref{Sz}) can be simplified to the following relations
\begin{eqnarray}
\frac{d^2 S_r}{d t^2}+\frac{1}{\tau}\frac{d S_r}{d t}
+\omega^2 S_r+\omega\Omega S_z=0,
\label{SrHomEq}\\
\frac{d^2 S_z}{d t^2}+\frac{1}{\tau}\frac{d S_z}{d t}
+\Omega^2 S_z+\omega\Omega S_r=0.
\label{SzHomEq}
\end{eqnarray}
Solving  these simple equations with the initial conditions (\ref{IC1}) and (\ref{IC2}), we find  that the relaxation of  homogeneous spin polarization initially directed along the ring's axis is described by
\begin{eqnarray}
S_r(t)=-\frac{\omega\Omega S_0}{\omega^2+\Omega^2}
\left[1- e^{-\frac{t}{2\tau}}\left(
\cosh\kappa t+
\frac{\sinh\kappa t}{2\tau\kappa}\right)\right], \;\;\;
\label{SrHom}\\
S_z(t)=\frac{\omega^2 S_0}{\omega^2+\Omega^2}
\left[1+\frac{\Omega^2}{\omega^2}e^{-\frac{t}{2\tau}}\left(
\cosh\kappa t+
\frac{\sinh\kappa t}{2\tau\kappa}\right)\right], \;\;\;
\label{SzHom}
\end{eqnarray}
where $\kappa=\sqrt{\frac{1}{4\tau^2}-(\omega^2+\Omega^2)}$. It should be emphasized that the parameter $\kappa$ can take both
real and imaginary values depending on system parameters. In particular, we can define a diffusive ($\tau\sqrt{\omega^2+\Omega^2}\ll 1$, $\textnormal{Im}(\kappa)=0$) and
ballistic ($\tau\sqrt{\omega^2+\Omega^2}\gg 1$, $\textnormal{Re}(\kappa)=0$) limits of spin dynamics. According to these definitions, in the diffusive limit the mean free path is much smaller than the ring radius, $\ell \ll r$, and the spin precession angle per mean free path is small, $\eta \ell \ll 1$. In the opposite ballistic limit, at least one of the following inequalities should hold: $\ell \gg r$, $\eta \ell \gg 1$. Any of these two conditions for the ballistic limit results in oscillations in spin relaxation dynamics. In the diffusive limit, however, the spin polarization decays exponentially without any oscillations.

\begin{figure} [tbp]
\begin{center}
\includegraphics[angle=0,width=8cm]{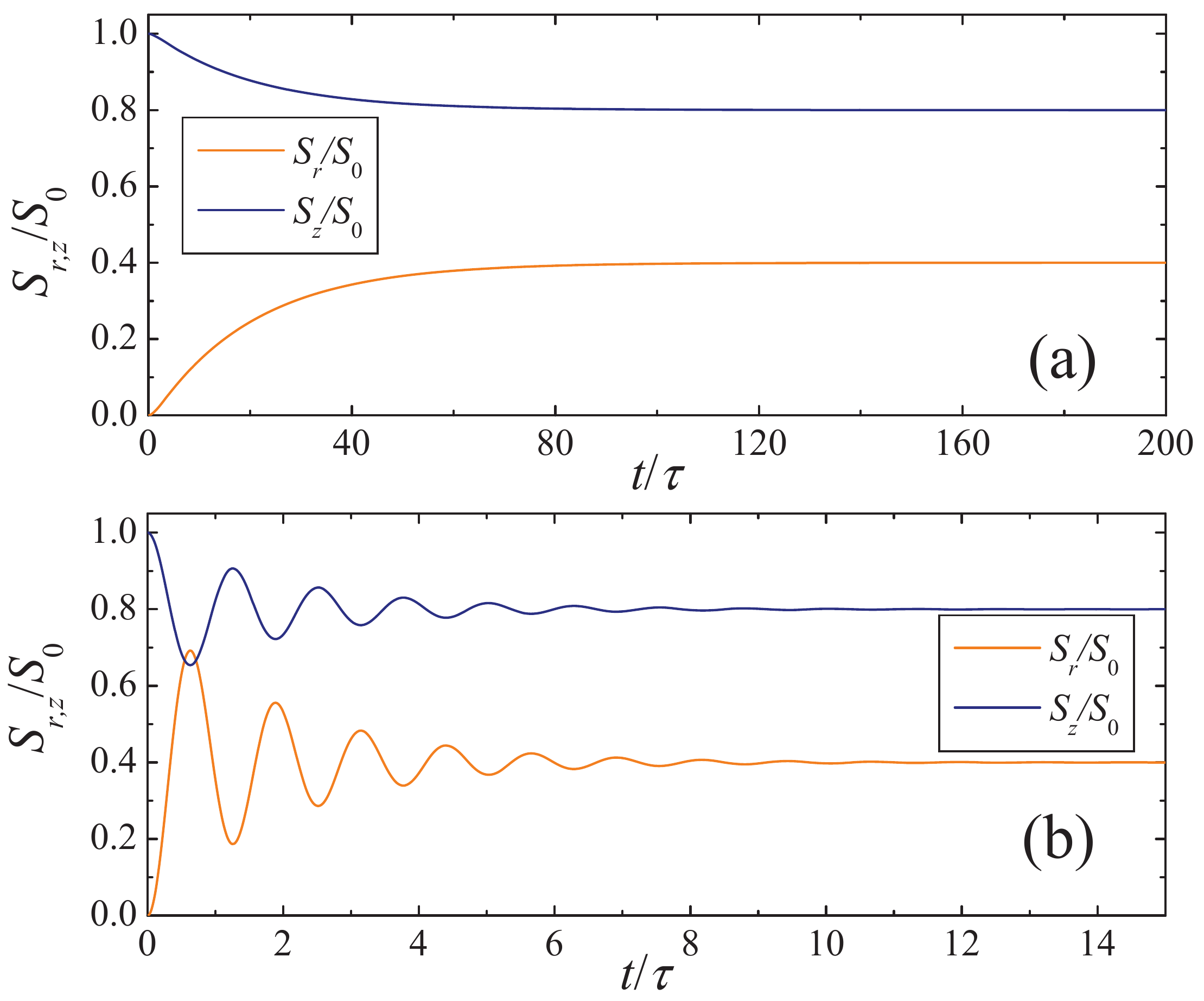}
\caption{(Color online) Relaxation of electron spin polarization in a ring in the (a) diffusive and (b) ballistic regimes. $S_\theta=0$.
This plot was obtained using Eqs. (\ref{SrHom}) and (\ref{SzHom}) with the following parameter values: $\Omega/\omega=-0.5$, $\tau\kappa=0.45$ (in (a)) and $\tau\kappa=i5$ (in (b)).}
\label{fig2}
\end{center}
\end{figure}

Eqs. (\ref{SrHom}) and (\ref{SzHom}) can be further simplified in the diffusive and ballistic limits. In diffusive limit, $\tau\sqrt{\omega^2+\Omega^2}\ll 1$, Eqs. (\ref{SrHom}) and (\ref{SzHom}) take the form
\begin{eqnarray}
S_r(t)=-\frac{\omega\Omega S_0}{\omega^2+\Omega^2}
\left[1- e^{-\tau (\omega^2+\Omega^2)t}\right],
\label{SrHom2}\\
S_z(t)=\frac{\omega^2 S_0}{\omega^2+\Omega^2}
\left[1+\frac{\Omega^2}{\omega^2}e^{-\tau (\omega^2+\Omega^2)t}\right].
\label{SzHom2}
\end{eqnarray}
It is clearly seen from the above expressions that in this case the relaxation of spin polarization is characterized by the time
$(\tau(\omega^2+\Omega^2))^{-1}$, which is much longer than $\tau$. In the opposite ballistic limit, when $\tau\sqrt{\omega^2+\Omega^2}\gg 1$, Eqs. (\ref{SrHom}) and (\ref{SzHom}) can be reduced to
\begin{eqnarray}
S_r(t)=-\frac{\omega\Omega S_0}{\omega^2+\Omega^2}
\left[1- e^{-\frac{t}{2\tau}}\cos(\sqrt{\omega^2+\Omega^2}t)\right],
\label{SrHom1}\\
S_z(t)=\frac{\omega^2 S_0}{\omega^2+\Omega^2}
\left[1+\frac{\Omega^2}{\omega^2}e^{-\frac{t}{2\tau}}\cos(\sqrt{\omega^2+\Omega^2}t)\right].
\label{SzHom1}
\end{eqnarray}
Consequently, the time-dependence of spin polarization components involves
an exponential decay with a time constant $2\tau$ modulated by oscillating functions.
Fig. \ref{fig2} shows an example of spin relaxation dynamics in both diffusive and ballistic limits.

The most important feature of the solution given by Eqs. (\ref{SrHom}) and (\ref{SzHom}) is that the electron spin polarization does not decay completely to zero. At long times, it shapes into a persistent crown-like spin polarization structure (schematically presented
in Fig. \ref{fig1}). The long-time values of spin polarization components are
\begin{eqnarray}
S_r=-\frac{\omega\Omega}{\omega^2+\Omega^2}S_0
, ~~
S_z=\frac{\omega^2 }{\omega^2+\Omega^2}S_0.
\label{Szstat}
\end{eqnarray}
The above relations indicate that during the spin relaxation process the amplitude of spin polarization decreases
by the factor
\begin{eqnarray}
\frac{S}{S_0}=\frac{\sqrt{S_r^2+S_z^2}}{S_0}=\frac{\omega}{\sqrt{\omega^2+\Omega^2}}.
\label{SzAttenuation}
\end{eqnarray}
In this persistent crown-like spin polarization
structure the angle between the ring's axis and the direction of spin polarization
does not depend on the magnitude of the initial polarization $S_0$ and is equal to
\begin{eqnarray}
\tan\psi=\frac{S_r}{S_z}=-\frac{\Omega}{\omega}=-\eta r.
\label{SzAngle}
\end{eqnarray}
It follows from Eqs. (\ref{SzAttenuation}) and (\ref{SzAngle}) that the geometric parameters of persistent crown-like
spin structure depend (besides the ring radius $r$) only on the ratio of the frequency of spin-orbit precession
to the frequency of the electron rotation, $\Omega / \omega$. Moreover, we note that in the case of  the opposite direction of the initial spin polarization  the asymptotic value of $S_r$ changes to the opposite one. In particular, if we consider the situation shown in Fig. \ref{fig1} (realizable at a negative value of $\alpha$), then the change $S_0\rightarrow -S_0$ will result in the opposite directions of the radial and $z$ components of spin polarization. Such an asymmetry can be considered as a result of the symmetry breaking by the spin-orbit interaction.

\section{Persistent spin states}
The crown-like persistent spin polarization structure discussed in the previous section is the simplest example of persistent spin states of spin kinetic equation (\ref{Sr})-(\ref{Sz}). In this section we find the complete set of the persistent spin states of Eqs. (\ref{Sr})-(\ref{Sz}). For this purpose, we search specific solutions of these equations in the complex form
\begin{eqnarray}
S(\theta,t)=\left(
   \begin{array}{c}
     S_r(\theta,t) \\
     S_\theta(\theta,t)\\
     S_z(\theta,t) \\
   \end{array}
 \right)=
 \left(
   \begin{array}{c}
     \tilde{S}_r \\
     \tilde{S}_\theta\\
     \tilde{S}_z \\
   \end{array}
 \right)e^{in\theta}e^{\lambda t},
\label{SAnsatz}
\end{eqnarray}
where $n$ is an integer number. The possible values of $\lambda$ and corresponding amplitudes $\tilde{S}_r$, $\tilde{S}_\theta$, $\tilde{S}_z$  have to be found. Substituting Eq. (\ref{SAnsatz}) into Eqs. (\ref{Sr})-(\ref{Sz}) we readily obtain a system of linear homogeneous equations.
The condition of consistency of these equations give rise for the following values of $\lambda$:
\begin{eqnarray}
\lambda_1^{\pm}&=&-\frac{1}{2\tau}\pm \sqrt{\frac{1}{4\tau^2}-n^2\omega^2},
\label{lambda1}\\
\lambda_2^{\pm}&=&-\frac{1}{2\tau}\pm \sqrt{\frac{1}{4\tau^2}-\left( n\omega+\sqrt{\omega^2+\Omega^2}\right)^2},
\label{lambda2}\\
\lambda_3^{\pm}&=&-\frac{1}{2\tau}\pm \sqrt{\frac{1}{4\tau^2}-\left(n\omega-\sqrt{\omega^2+\Omega^2}\right)^2}.
\label{lambda3}
\end{eqnarray}
Solving the linear homogeneous equations for each value of $\lambda$ (given by Eqs. (\ref{lambda1})-(\ref{lambda3})) we find corresponding amplitudes
 \begin{eqnarray}
 \tilde{S}_1=\left(
   \begin{array}{c}
     -\frac{\Omega}{\sqrt{\omega^2+\Omega^2}} \\
    0 \\
     \frac{\omega}{\sqrt{\omega^2+\Omega^2}} \\
   \end{array}
 \right),
 \tilde{S}_2=\frac{1}{\sqrt{2}}\left(
   \begin{array}{c}
     \frac{\omega}{\sqrt{\omega^2+\Omega^2}} \\
     -i \\
     \frac{\Omega}{\sqrt{\omega^2+\Omega^2}} \\
   \end{array}
 \right),\nonumber\\
 \tilde{S}_3=\frac{1}{\sqrt{2}}\left(
   \begin{array}{c}
     \frac{\omega}{\sqrt{\omega^2+\Omega^2}} \\
     i \\
     \frac{\Omega}{\sqrt{\omega^2+\Omega^2}} \\
   \end{array}
 \right).~~~~~~~~~~~~~~~
\label{EigenVectors}
\end{eqnarray}
It is interesting to note that the vectors given by Eqs. (\ref{EigenVectors}) do not depend on $\tau$ and $n$. They are orthogonal and normalized (in complex sense), so it is easy to expand any vector (for example, related to initial spin polarization) in this basis.

A spin state is persistent when $\lambda=0$ (this follows directly from Eq. (\ref{SAnsatz})).
It can be seen from Eqs. (\ref{lambda1})-(\ref{lambda3}) that there are three possibilities when the condition $\lambda=0$ is realized.
The first possibility is when $n=0$, in this case, $\lambda^{-}_1$ is equal to zero.
The corresponding persistent spin polarization structure is given by $S=A\tilde{S}_1$, that is
\begin{eqnarray}
S_r=-\frac{\Omega}{\sqrt{\omega^2+\Omega^2}} A,~S_\theta=0,~S_z=\frac{\omega}{\sqrt{\omega^2+\Omega^2}} A,
\label{SpinCrown}
\end{eqnarray}
where $A$ is an arbitrary constant describing the magnitude of spin polarization in this state.
It can be seen that the persistent spin state given by Eq. (\ref{SpinCrown}) is realized in the relaxation
of homogeneous spin polarization initially directed in $z$ direction (this problem was considered in the previous section, see Eq. (\ref{Szstat})).

\begin{figure}
\begin{center}
\includegraphics[angle=0,width=4.0cm]{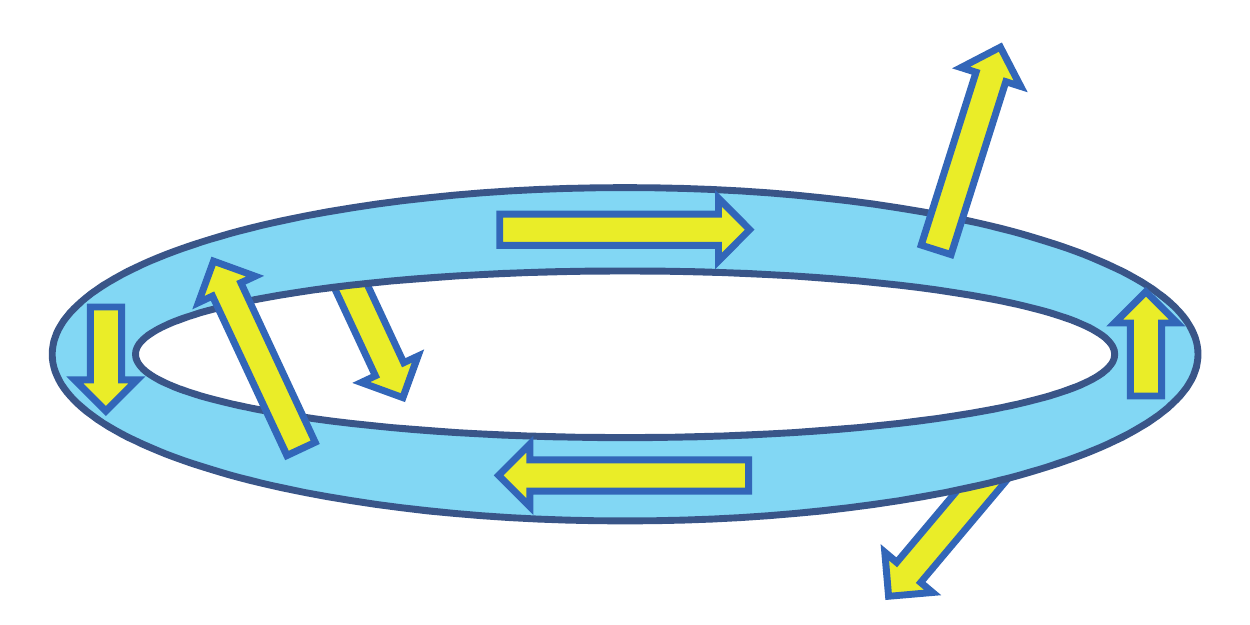}
\caption{(Color online) $n=2$ stationary state (as follows from Eq. (\ref{SpinExotic}) at $\theta_0=0$).}
\label{fig3}
\end{center}
\end{figure}

As it is seen from Eqs. (\ref{lambda2}) and (\ref{lambda3}), two other possibilities ($\lambda^{-}_{2}=0$ or $\lambda^{-}_{3}=0$) may only be realized  if $n=-\sqrt{1+\frac{\Omega^2}{\omega^2}}$ or $n=\sqrt{1+\frac{\Omega^2}{\omega^2}}$ correspondingly. Since $n$ is an integer number, these conditions are fulfilled only for certain combinations of system parameters as discussed below.  By substituting Eqs. (\ref{EigenVectors}) for $\tilde{S}_2$ or $\tilde{S}_3$ into Eq. (\ref{SAnsatz}) and taking real or imaginary parts, we obtain the persistent spin polarization structures that can be presented as
\begin{eqnarray}
S_r=\frac{B}{n} \sin n(\theta-\theta_0),S_\theta=B\cos n(\theta-\theta_0),\nonumber\\
S_z= \frac{B\sqrt{n^2-1}}{n}\textnormal{sgn}\left( \Omega\right)\sin n(\theta-\theta_0),
\label{SpinExotic}
\end{eqnarray}
where $n=\sqrt{1+\frac{\Omega^2}{\omega^2}}=2,3,4,...$, $B$ and $\theta_0$ are arbitrary real constants, which determine spin polarization amplitude and angular position of the structure respectively. Fig. \ref{fig3} shows schematically the persistent spin polarization structure (\ref{SpinExotic}) for $n=2$.

It should be noted that the persistent spin polarization structures (Eq. (\ref{SpinExotic})) exist only at certain relations between the ring's radius $r$ and the strength of spin-orbit interaction $\eta$. Since $\Omega/\omega=\eta r$,
the conditions for persistent spin states are given by
\begin{eqnarray}
|\eta| r=\sqrt{n^2-1},~~ n=2,3,4,...
\label{SpinExoticParameters}
\end{eqnarray}
In addition, we note that at $r\rightarrow \infty$, $n$ becomes much greater than $1$. In this case in accordance with Eq. (\ref{SpinExoticParameters})
$n\approx \eta r\gg 1 $, and persistent spin polarization structure (\ref{SpinExotic}) becomes the usual plain spin helix \cite{Pershin05a} with wave vector $\eta$.

\section{Green function}
Linear combinations of specific solutions (\ref{SAnsatz}) (with amplitudes (\ref{EigenVectors}) and corresponding $\lambda$-s given by Eqs. (\ref{lambda1})-(\ref{lambda3})) of homogeneous linear Eqs. (\ref{Sr})-(\ref{Sz}) are also
solutions of these equations. Therefore, the general solution of the system (\ref{Sr})-(\ref{Sz}) can be presented as a sum
\begin{eqnarray}
S(\theta,t)=\sum_{n,\nu,\sigma}A_\nu^\sigma(n)\tilde{S}_\nu e^{\lambda^\sigma_\nu(n) t}e^{in\theta},
\label{GenSol1}
\end{eqnarray}
where  $A_\nu^\sigma(n)$ are complex constants, $\nu=1,2,3$,  $\sigma=+,-$, and $n$ runs over all integer values.
In order to find the constants $A_\nu^\sigma(n)$ we should specify the initial conditions for
spin polarization
\begin{eqnarray}
S(\theta,t)_{t=0}\equiv S(\theta,0),
\left(\frac{\partial S(\theta,t)}{\partial t}\right)_{t=0}\equiv\dot{S}(\theta,0).
\label{InitCond}
\end{eqnarray}
Using initial conditions (\ref{InitCond}) we  obtain
\begin{eqnarray}
A_\nu^\pm (n)=\int_{0}^{2\pi}\frac{d\theta_0}{2\pi}e^{-in\theta_0}
\frac{\langle\tilde{S}_\nu,\dot{S}(\theta_0,0)-\lambda^\mp_\nu(n)S(\theta_0,0)\rangle}{\lambda^\pm_\nu(n)-\lambda^\mp_\nu(n)}, \;\;
\label{An}
\end{eqnarray}
where $\langle\phi,\psi\rangle=\overline{\phi}_r\psi_r+\overline{\phi}_\theta\psi_\theta+\overline{\phi}_z\psi_z$ is the inner product of amplitudes. It is easy to check that
$\langle\tilde{S}_\nu,\tilde{S}_\mu\rangle=\delta_{\nu\mu}$.

The Green function of spin kinetic equation can be obtained substituting Eq. (\ref{An}) into Eq. (\ref{GenSol1}).
The Green function can be employed to find spin polarization at any moment of time for any given initial conditions as
\begin{eqnarray}
S_{\alpha}(\theta,t)=\left[\frac{\partial}{\partial t}+\frac{1}{\tau} \right]\int_{0}^{2\pi}d\theta_0 G_{\alpha\beta}(\theta-\theta_0,t)S_\beta(\theta_0,0)\nonumber\\
+\int_{0}^{2\pi}d\theta_0 G_{\alpha\beta}(\theta-\theta_0,t)\dot{S}_\beta(\theta_0,0), \;\;\;
\label{GenSol2}
\end{eqnarray}
where $\alpha$ or $\beta$ take the values $r,\theta,z$ and summation over repeated indexes is implied.

In particular, the components of the Green function (it is actually 3x3 matrix) of the system (\ref{Sr})-(\ref{Sz}) can be written as
\begin{eqnarray}
G_{rr}(\theta,t)=\frac{\Omega^2 G_0(\theta,t)+\omega^2G_c(\theta,t)}{\omega^2+\Omega^2},
\label{Grr}
\end{eqnarray}
\begin{eqnarray}
G_{\theta r}(\theta,t)=-\frac{\omega G_s(\theta,t)}{\sqrt{\omega^2+\Omega^2}},
\label{Gtr}
\end{eqnarray}
\begin{eqnarray}
G_{zr}(\theta,t)=-\frac{\omega\Omega [G_0(\theta,t)-G_c(\theta,t)]}{\omega^2+\Omega^2},
\label{Gzr}
\end{eqnarray}

\begin{eqnarray}
G_{r\theta}(\theta,t)=-G_{\theta r}(\theta,t)
\label{Grt}
\end{eqnarray}
\begin{eqnarray}
G_{\theta \theta}(\theta,t)=G_{c}(\theta,t),
\label{Gtt}
\end{eqnarray}
\begin{eqnarray}
G_{z\theta}(\theta,t)=\frac{\Omega G_s(\theta,t)}{\sqrt{\omega^2+\Omega^2}},
\label{Gzt}
\end{eqnarray}

\begin{eqnarray}
G_{rz}(\theta,t)=G_{zr}(\theta,t)
\label{Grz}
\end{eqnarray}
\begin{eqnarray}
G_{\theta z}(\theta,t)=-G_{z\theta }(\theta,t),
\label{Gtz}
\end{eqnarray}
\begin{eqnarray}
G_{zz}(\theta,t)=\frac{\omega^2 G_0(\theta,t)+\Omega^2G_c(\theta,t)}{\omega^2+\Omega^2},
\label{Gzz}
\end{eqnarray}

where
\begin{eqnarray}
G_0(\theta,t)=\frac{e^{-\frac{t}{2\tau}}}{2\pi}\sum_{n=-\infty}^{n=+\infty}
\cos n\theta\frac{\sin\left(t\sqrt{\omega^2 n^2-1/(2\tau)^2}\right)}{\sqrt{\omega^2 n^2-1/(2\tau)^2}}, \;\;\;
\label{G0}
\end{eqnarray}
\begin{eqnarray}
G_c(\theta,t)=\frac{e^{-\frac{t}{2\tau}}}{4\pi}\sum_{n=-\infty}^{n=+\infty}
\cos n\theta
\left\{
\frac{\sin\Omega^{+}_n t}{\Omega^{+}_n}
+
\frac{\sin\Omega^{-}_n t}
{\Omega^{-}_n}
\right\},\;\;\;
\label{Gc}
\end{eqnarray}
\begin{eqnarray}
G_s(\theta,t)=-\frac{e^{-\frac{t}{2\tau}}}{4\pi}\sum_{n=-\infty}^{n=+\infty}
\sin n\theta
\left\{
\frac{\sin\Omega^{+}_n t}{\Omega^{+}_n}
-
\frac{\sin\Omega^{-}_n t}
{\Omega^{-}_n}
\right\},\;\;\;
\label{Gs}
\end{eqnarray}
\begin{eqnarray}
\Omega^{\pm}_n=\sqrt{\left(\omega n\pm\sqrt{\omega^2+\Omega^2}\right)^2-1/(2\tau)^2}.
\label{OmegaPM}
\end{eqnarray}

\begin{figure} [tbp]
\begin{center}
\includegraphics[angle=0,width=8cm]{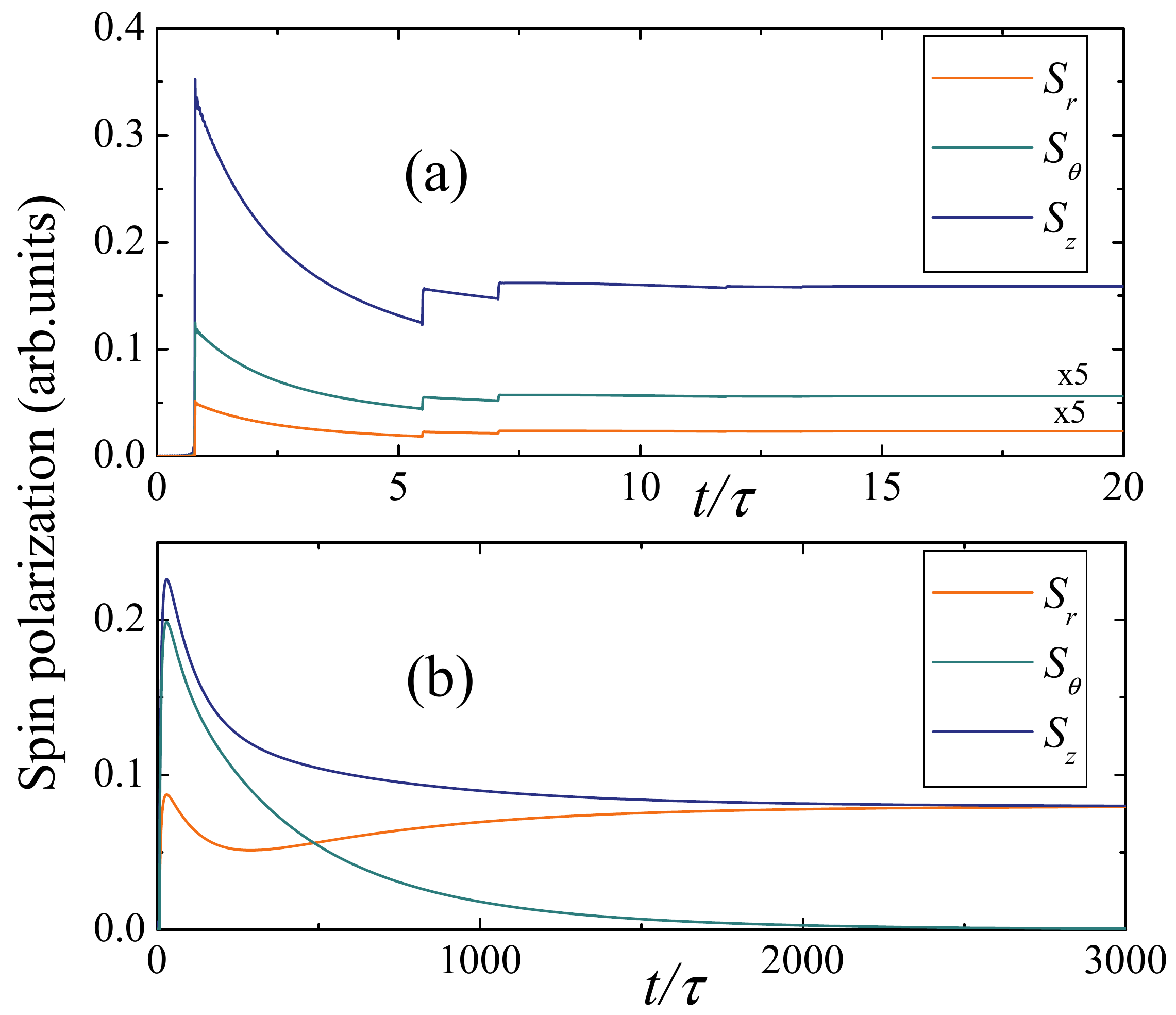}
\caption{(Color online) Spin polarization components at $\theta=\pi /4$ induced by a delta function excitation at $\theta=0$ ($S_\beta(\theta,t=0)=0$, $\dot{S}_\beta(\theta,t=0)=\delta(\theta)\delta_{\beta,z}$, where $\delta(..)$ is the delta function and $\delta_{i,j}$ is the Kronecker delta). This plot was obtained using the parameter values $\tau \Omega=-0.1$, $\tau \omega=1$ (in (a)), and $\tau \omega=0.1$ (in (b)).}
\label{fig4}
\end{center}
\end{figure}

The above equations can be further simplified. In fact, the sums appearing in Eqs. (\ref{G0})-(\ref{Gs}) can be calculated in the following way.
First of all, we employ the well-known formula from the Bessel function theory
\begin{eqnarray}
\frac{\sin \left(t\sqrt{\omega^2-q^2} \right)}{\sqrt{\omega^2-q^2}}= \nonumber \\
\frac{1}{2}
\int_{-\infty}^{+\infty}d\xi e^{i\omega\xi}\Phi(t-|\xi|)I_0\left( q\sqrt{t^2-\xi^2}\right)
\label{Formula},
\end{eqnarray}
where $\Phi(t)$ is the Heaviside step function, and $I_0(t)$ is the modified Bessel function
of zero order. Substituting Eq. (\ref{Formula}) into Eqs. (\ref{G0})-(\ref{Gs}) we perform summation in
these equations taking into account that \cite{VladimirovEqMathPhys}
\begin{eqnarray}
\sum_{n=-\infty}^{n=+\infty}e^{in\xi}=2\pi\sum_{n=-\infty}^{n=+\infty}\delta(\xi-2\pi n).
\label{Formula2}
\end{eqnarray}
Then, the integrals in Eqs. (\ref{G0})-(\ref{Gs}) modified by Eq. (\ref{Formula}) are trivially evaluated yielding the following results

\begin{eqnarray}
G_0(\theta,t)=\frac{e^{-\frac{t}{2\tau}}}{2\omega}\sum_{n=-\infty}^{n=+\infty}
I_0
\left(
\frac{\sqrt{\omega^2t^2-(\theta-2\pi n)^2}}{2\omega\tau}
\right)\nonumber\\
\times\Phi(\omega t-|\theta-2\pi n|),~
\label{G02}
\end{eqnarray}

\begin{eqnarray}
G_c(\theta,t)=\frac{e^{-\frac{t}{2\tau}}}{2\omega}\sum_{n=-\infty}^{n=+\infty}
I_0
\left(
\frac{\sqrt{\omega^2t^2-(\theta-2\pi n)^2}}{2\omega\tau}
\right)\nonumber\\
\times\cos\left[
\sqrt{1+\frac{\Omega^2}{\omega^2}}\left(\theta-2\pi n\right)\right]
\Phi(\omega t-|\theta-2\pi n|),
\label{Gc2}
\end{eqnarray}

\begin{eqnarray}
G_s(\theta,t)=\frac{e^{-\frac{t}{2\tau}}}{2\omega}\sum_{n=-\infty}^{n=+\infty}
I_0
\left(
\frac{\sqrt{\omega^2t^2-(\theta-2\pi n)^2}}{2\omega\tau}
\right)\nonumber\\
\times\sin\left[
\sqrt{1+\frac{\Omega^2}{\omega^2}}\left(\theta-2\pi n\right)\right]
\Phi(\omega t-|\theta-2\pi n|).
\label{Gs2}
\end{eqnarray}

In fact, all sums in Eqs. (\ref{G02})-(\ref{Gs2}) are finite sums because of the Heaviside functions $\Phi(\omega t-|\theta-2\pi n|)$.
For any given values of time $t$ and angle $\theta$ only the terms with $n$ in the range  $(\theta-\omega t)/(2\pi)<n<(\theta+\omega t)/(2\pi)$
give non-zero contribution to the R.H.S. of Eqs. (\ref{G02})-(\ref{Gs2}).

Fig. \ref{fig4} shows an example of application of the Green function formalism to calculations of spin polarization dynamics.
Specifically, we plot spin polarization components at $\theta=\pi /4$ induced by spin polarization excitation at $\theta=0$. Such an excitation
induces clockwise and counterclockwise propagating waves of spin polarization that finally evolve into a steady homogeneous spin polarization.
Each time when clockwise and counterclockwise propagating waves of spin polarization reach $\theta=\pi /4$, the spin polarization at this point changes in steps (see Fig. \ref{fig4}(a)). The step amplitudes, however, decrease in time because of the diffusive component in spin transport. This type of behavior resembles a multiple echo.
In addition, in the diffusive limit shown \ref{fig4}(b) we did not observe clear steps in spin polarization. At long times, $S_\theta \rightarrow 0$ and the crown-like spin polarization configuration is formed.

\section{Conclusions}

Spin relaxation in confined structures exhibits some remarkable properties and behavior that are not found in infinite 2D systems.
In this paper we have studied the electron spin relaxation in the ring. It has been found that the homogeneous spin polarization along the ring axis
transforms into a persistent crown-like spin structure. Moreover, a family of persistent spin states in the ring has been identified. We have also derived the Green function
of spin kinetic equation and used it to investigate the propagation of a point spin excitation.

\bibliography{spin}

\begin{thebibliography}{29}
\expandafter\ifx\csname natexlab\endcsname\relax\def\natexlab#1{#1}\fi
\expandafter\ifx\csname bibnamefont\endcsname\relax
  \def\bibnamefont#1{#1}\fi
\expandafter\ifx\csname bibfnamefont\endcsname\relax
  \def\bibfnamefont#1{#1}\fi
\expandafter\ifx\csname citenamefont\endcsname\relax
  \def\citenamefont#1{#1}\fi
\expandafter\ifx\csname url\endcsname\relax
  \def\url#1{\texttt{#1}}\fi
\expandafter\ifx\csname urlprefix\endcsname\relax\def\urlprefix{URL }\fi
\providecommand{\bibinfo}[2]{#2}
\providecommand{\eprint}[2][]{\url{#2}}

\bibitem[{\citenamefont{Zutic et~al.}(2004)\citenamefont{Zutic, Fabian, and
  Das~Sarma}}]{Zutic04a}
\bibinfo{author}{\bibfnamefont{I.}~\bibnamefont{Zutic}},
  \bibinfo{author}{\bibfnamefont{J.}~\bibnamefont{Fabian}}, \bibnamefont{and}
  \bibinfo{author}{\bibfnamefont{S.}~\bibnamefont{Das~Sarma}},
  \bibinfo{journal}{Rev. Mod. Phys.} \textbf{\bibinfo{volume}{76}},
  \bibinfo{pages}{323} (\bibinfo{year}{2004}).

\bibitem[{\citenamefont{Wu et~al.}(2010)\citenamefont{Wu, Jiang, and
  Weng}}]{Wu10a}
\bibinfo{author}{\bibfnamefont{M.}~\bibnamefont{Wu}},
  \bibinfo{author}{\bibfnamefont{J.}~\bibnamefont{Jiang}}, \bibnamefont{and}
  \bibinfo{author}{\bibfnamefont{M.}~\bibnamefont{Weng}},
  \bibinfo{journal}{Phys. Reports} \textbf{\bibinfo{volume}{439}},
  \bibinfo{pages}{61} (\bibinfo{year}{2010}).

\bibitem[{\citenamefont{Dyakonov and {Perel'}}(1972)}]{Dyakonov72a}
\bibinfo{author}{\bibfnamefont{M.~I.} \bibnamefont{Dyakonov}} \bibnamefont{and}
  \bibinfo{author}{\bibfnamefont{V.~I.} \bibnamefont{{Perel'}}},
  \bibinfo{journal}{Sov. Phys. Solid State} \textbf{\bibinfo{volume}{13}},
  \bibinfo{pages}{3023} (\bibinfo{year}{1972}).

\bibitem[{\citenamefont{Dyakonov and Kachorovskii}(1986)}]{Dyakonov86a}
\bibinfo{author}{\bibfnamefont{M.~I.} \bibnamefont{Dyakonov}} \bibnamefont{and}
  \bibinfo{author}{\bibfnamefont{V.~Y.} \bibnamefont{Kachorovskii}},
  \bibinfo{journal}{Sov. Phys. Semicond.} \textbf{\bibinfo{volume}{20}},
  \bibinfo{pages}{110} (\bibinfo{year}{1986}).

\bibitem[{\citenamefont{Sherman}(2003)}]{Sherman03a}
\bibinfo{author}{\bibfnamefont{E.~Y.} \bibnamefont{Sherman}},
  \bibinfo{journal}{Appl. Phys. lett} \textbf{\bibinfo{volume}{82}},
  \bibinfo{pages}{209} (\bibinfo{year}{2003}).

\bibitem[{\citenamefont{Pershin}(2005{\natexlab{a}})}]{Pershin05a}
\bibinfo{author}{\bibfnamefont{Y.~V.} \bibnamefont{Pershin}},
  \bibinfo{journal}{Phys. Rev. B} \textbf{\bibinfo{volume}{71}},
  \bibinfo{pages}{155317} (\bibinfo{year}{2005}{\natexlab{a}}).

\bibitem[{\citenamefont{Bernevig et~al.}(2006)\citenamefont{Bernevig,
  Orenstein, and Zhang}}]{Bernevig06a}
\bibinfo{author}{\bibfnamefont{B.~A.} \bibnamefont{Bernevig}},
  \bibinfo{author}{\bibfnamefont{J.}~\bibnamefont{Orenstein}},
  \bibnamefont{and} \bibinfo{author}{\bibfnamefont{S.-C.} \bibnamefont{Zhang}},
  \bibinfo{journal}{Phys. Rev. Lett.} \textbf{\bibinfo{volume}{97}},
  \bibinfo{pages}{236601} (\bibinfo{year}{2006}).

\bibitem[{\citenamefont{Weng et~al.}(2008)\citenamefont{Weng, Wu, and
  Cui}}]{Weng08a}
\bibinfo{author}{\bibfnamefont{M.~Q.} \bibnamefont{Weng}},
  \bibinfo{author}{\bibfnamefont{M.~W.} \bibnamefont{Wu}}, \bibnamefont{and}
  \bibinfo{author}{\bibfnamefont{H.~L.} \bibnamefont{Cui}},
  \bibinfo{journal}{J. Appl. Phys.} \textbf{\bibinfo{volume}{103}},
  \bibinfo{pages}{063714} (\bibinfo{year}{2008}).

\bibitem[{\citenamefont{Kleinert and Bryksin}(2009)}]{Kleinert09a}
\bibinfo{author}{\bibfnamefont{P.}~\bibnamefont{Kleinert}} \bibnamefont{and}
  \bibinfo{author}{\bibfnamefont{V.~V.} \bibnamefont{Bryksin}},
  \bibinfo{journal}{Phys. Rev. B} \textbf{\bibinfo{volume}{79}},
  \bibinfo{pages}{045317} (\bibinfo{year}{2009}).

\bibitem[{\citenamefont{Tokatly and Sherman}(2010)}]{Tokatly10a}
\bibinfo{author}{\bibfnamefont{I.~V.} \bibnamefont{Tokatly}} \bibnamefont{and}
  \bibinfo{author}{\bibfnamefont{E.~Y.} \bibnamefont{Sherman}},
  \bibinfo{journal}{Ann. Phys.} \textbf{\bibinfo{volume}{325}},
  \bibinfo{pages}{1104} (\bibinfo{year}{2010}).

\bibitem[{\citenamefont{Pershin and Slipko}(2010{\natexlab{a}})}]{pershin10a}
\bibinfo{author}{\bibfnamefont{Y.~V.} \bibnamefont{Pershin}} \bibnamefont{and}
  \bibinfo{author}{\bibfnamefont{V.~A.} \bibnamefont{Slipko}},
  \bibinfo{journal}{Phys. Rev. B} \textbf{\bibinfo{volume}{82}},
  \bibinfo{pages}{125325} (\bibinfo{year}{2010}{\natexlab{a}}).

\bibitem[{\citenamefont{Galitski et~al.}(2006)\citenamefont{Galitski, Burkov,
  and Das~Sarma}}]{Galitski06a}
\bibinfo{author}{\bibfnamefont{V.~M.} \bibnamefont{Galitski}},
  \bibinfo{author}{\bibfnamefont{A.~A.} \bibnamefont{Burkov}},
  \bibnamefont{and}
  \bibinfo{author}{\bibfnamefont{S.}~\bibnamefont{Das~Sarma}},
  \bibinfo{journal}{Phys. Rev. B} \textbf{\bibinfo{volume}{74}},
  \bibinfo{pages}{115331} (\bibinfo{year}{2006}).

\bibitem[{\citenamefont{Kiselev and Kim}(2000)}]{Kiselev00a}
\bibinfo{author}{\bibfnamefont{A.~A.} \bibnamefont{Kiselev}} \bibnamefont{and}
  \bibinfo{author}{\bibfnamefont{K.~W.} \bibnamefont{Kim}},
  \bibinfo{journal}{Phys. Rev. B} \textbf{\bibinfo{volume}{61}},
  \bibinfo{pages}{13115} (\bibinfo{year}{2000}).

\bibitem[{\citenamefont{Holleitner et~al.}(2007)\citenamefont{Holleitner, Sih,
  Myers, Gossard, and Awschalom}}]{Holleitner07a}
\bibinfo{author}{\bibfnamefont{A.~W.} \bibnamefont{Holleitner}},
  \bibinfo{author}{\bibfnamefont{V.}~\bibnamefont{Sih}},
  \bibinfo{author}{\bibfnamefont{R.~C.} \bibnamefont{Myers}},
  \bibinfo{author}{\bibfnamefont{A.~C.} \bibnamefont{Gossard}},
  \bibnamefont{and} \bibinfo{author}{\bibfnamefont{D.~D.}
  \bibnamefont{Awschalom}}, \bibinfo{journal}{New J. Phys.}
  \textbf{\bibinfo{volume}{9}}, \bibinfo{pages}{342} (\bibinfo{year}{2007}).

\bibitem[{\citenamefont{Pershin}(2005{\natexlab{b}})}]{Pershin05c}
\bibinfo{author}{\bibfnamefont{Y.~V.} \bibnamefont{Pershin}},
  \bibinfo{journal}{Phys. E} \textbf{\bibinfo{volume}{27}}, \bibinfo{pages}{77}
  (\bibinfo{year}{2005}{\natexlab{b}}).

\bibitem[{\citenamefont{Pershin and Privman}(2004)}]{Pershin04a}
\bibinfo{author}{\bibfnamefont{Y.~V.} \bibnamefont{Pershin}} \bibnamefont{and}
  \bibinfo{author}{\bibfnamefont{V.}~\bibnamefont{Privman}},
  \bibinfo{journal}{Phys. Rev. B} \textbf{\bibinfo{volume}{69}},
  \bibinfo{pages}{073310} (\bibinfo{year}{2004}).

\bibitem[{\citenamefont{Lyubinskiy}(2006)}]{Lyubinskiy06a}
\bibinfo{author}{\bibfnamefont{I.~S.} \bibnamefont{Lyubinskiy}},
  \bibinfo{journal}{JETP Lett.} \textbf{\bibinfo{volume}{83}},
  \bibinfo{pages}{336} (\bibinfo{year}{2006}).

\bibitem[{\citenamefont{Slipko and Pershin}(2011)}]{Slipko11a}
\bibinfo{author}{\bibfnamefont{V.~A.} \bibnamefont{Slipko}} \bibnamefont{and}
  \bibinfo{author}{\bibfnamefont{Y.~V.} \bibnamefont{Pershin}},
  \bibinfo{journal}{arXiv:1102.0039}  (\bibinfo{year}{2011}).

\bibitem[{\citenamefont{Bychkov and Rashba}(1984)}]{Bychkov84a}
\bibinfo{author}{\bibfnamefont{Y.}~\bibnamefont{Bychkov}} \bibnamefont{and}
  \bibinfo{author}{\bibfnamefont{E.}~\bibnamefont{Rashba}},
  \bibinfo{journal}{{JETP} Lett.} \textbf{\bibinfo{volume}{39}},
  \bibinfo{pages}{78} (\bibinfo{year}{1984}).

\bibitem[{\citenamefont{Yu and Flatt\'e}(2002)}]{Yu02a}
\bibinfo{author}{\bibfnamefont{Z.~G.} \bibnamefont{Yu}} \bibnamefont{and}
  \bibinfo{author}{\bibfnamefont{M.~E.} \bibnamefont{Flatt\'e}},
  \bibinfo{journal}{Phys. Rev. B} \textbf{\bibinfo{volume}{66}},
  \bibinfo{pages}{201202} (\bibinfo{year}{2002}).

\bibitem[{\citenamefont{Burkov et~al.}(2004)\citenamefont{Burkov, N\'u\~nez,
  and MacDonald}}]{Burkov04a}
\bibinfo{author}{\bibfnamefont{A.~A.} \bibnamefont{Burkov}},
  \bibinfo{author}{\bibfnamefont{A.~S.} \bibnamefont{N\'u\~nez}},
  \bibnamefont{and} \bibinfo{author}{\bibfnamefont{A.~H.}
  \bibnamefont{MacDonald}}, \bibinfo{journal}{Phys. Rev. B}
  \textbf{\bibinfo{volume}{70}}, \bibinfo{pages}{155308}
  (\bibinfo{year}{2004}).

\bibitem[{\citenamefont{Saikin}(2004)}]{Saikin04a}
\bibinfo{author}{\bibfnamefont{S.}~\bibnamefont{Saikin}}, \bibinfo{journal}{J.
  Phys.: Condens. Matter} \textbf{\bibinfo{volume}{16}}, \bibinfo{pages}{5071}
  (\bibinfo{year}{2004}).

\bibitem[{\citenamefont{Pershin}(2004)}]{Pershin04b}
\bibinfo{author}{\bibfnamefont{Y.~V.} \bibnamefont{Pershin}},
  \bibinfo{journal}{Phys. E} \textbf{\bibinfo{volume}{23}},
  \bibinfo{pages}{226} (\bibinfo{year}{2004}).

\bibitem[{\citenamefont{Pershin and
  Slipko}(2010{\natexlab{b}})}]{arXiv_1007_0853v1}
\bibinfo{author}{\bibfnamefont{Y.~V.} \bibnamefont{Pershin}} \bibnamefont{and}
  \bibinfo{author}{\bibfnamefont{V.~A.} \bibnamefont{Slipko}},
  \bibinfo{journal}{arXiv:1007.0853v1}  (\bibinfo{year}{2010}{\natexlab{b}}).

\bibitem[{\citenamefont{Meijer et~al.}(2002)\citenamefont{Meijer, Morpurgo, and
  Klapwijk}}]{Meijer02a}
\bibinfo{author}{\bibfnamefont{F.~E.} \bibnamefont{Meijer}},
  \bibinfo{author}{\bibfnamefont{A.~F.} \bibnamefont{Morpurgo}},
  \bibnamefont{and} \bibinfo{author}{\bibfnamefont{T.~M.}
  \bibnamefont{Klapwijk}}, \bibinfo{journal}{Phys. Rev. B}
  \textbf{\bibinfo{volume}{66}}, \bibinfo{pages}{033107}
  (\bibinfo{year}{2002}).

\bibitem[{\citenamefont{Berche et~al.}(2010)\citenamefont{Berche, Chatelain,
  and Medina}}]{Berche10a}
\bibinfo{author}{\bibfnamefont{B.}~\bibnamefont{Berche}},
  \bibinfo{author}{\bibfnamefont{C.}~\bibnamefont{Chatelain}},
  \bibnamefont{and} \bibinfo{author}{\bibfnamefont{E.}~\bibnamefont{Medina}},
  \bibinfo{journal}{Eur. J. Phys.} \textbf{\bibinfo{volume}{31}},
  \bibinfo{pages}{1267} (\bibinfo{year}{2010}).

\bibitem[{\citenamefont{Lyubinskiy and Kachorovskii}(2006)}]{Lyubinskiy06b}
\bibinfo{author}{\bibfnamefont{I.~S.} \bibnamefont{Lyubinskiy}}
  \bibnamefont{and} \bibinfo{author}{\bibfnamefont{V.~Y.}
  \bibnamefont{Kachorovskii}}, \bibinfo{journal}{Phys. Rev. B}
  \textbf{\bibinfo{volume}{73}}, \bibinfo{pages}{041301}
  (\bibinfo{year}{2006}).

\bibitem[{\citenamefont{Pitaevskii and Lifshitz}(1981)}]{Pitaevskii81a}
\bibinfo{author}{\bibfnamefont{L.~P.} \bibnamefont{Pitaevskii}}
  \bibnamefont{and} \bibinfo{author}{\bibfnamefont{E.~M.}
  \bibnamefont{Lifshitz}}, \emph{\bibinfo{title}{Physical Kinetics}}
  (\bibinfo{publisher}{Butterworth-Heinemann}, \bibinfo{year}{1981}).

\bibitem[{\citenamefont{Vladimirov}(1985)}]{VladimirovEqMathPhys}
\bibinfo{author}{\bibfnamefont{V.}~\bibnamefont{Vladimirov}},
  \emph{\bibinfo{title}{Equations of Mathematical Physics}}
  (\bibinfo{publisher}{Mir}, \bibinfo{year}{1985}).

\end{thebibliography}

\end{document}